\documentclass[twocolumn,epsf,showpacs,prd]{revtex4}
\usepackage{graphicx}
\usepackage{dcolumn}
\usepackage{bm}
\begin{document}

\newcommand{\be}{\begin{equation}}
\newcommand{\ee}{\end{equation}}
\newcommand{\ba}{\begin{eqnarray}}
\newcommand{\ea}{\end{eqnarray}}
\def\bone{B^{(1)}}
\title{Gamma rays from the Galactic bulge and large extra dimensions}

\author{Michel Cass\'{e}$^{a,b}$, Jacques Paul$^{a,c}\footnote{Correspondence and requests for materials should be addressed to J.P. (e-mail: jpaul@cea.fr).
}$, Gianfranco Bertone$^{b}$, \& G\"unter Sigl$^{b,c}$}

\affiliation{$^a$ DAPNIA/Service d'Astrophysique, CEA-Saclay, 91191 
Gif-sur-Yvette Cedex, France \\
$^b$ Institut d'Astrophysique de Paris, 98bis, Boulevard Arago, 
75014 Paris, France \\
$^c$ F\'{e}d\'{e}ration de Recherche Astroparticule et Cosmologie, 
Universit\'{e} Paris 7, 2 place Jussieu, 75251 Paris Cedex 05, France \\}


\begin{abstract}

An intriguing feature of extra dimensions
is the possible production of Kaluza--Klein gravitons by nucleon-nucleon
bremsstrahlung in the course of core collapse of massive stars. In this
event Kaluza--Klein gravitons are copiously produced and a significant
fraction of them remains trapped around the newly born
neutron stars. They slowly decay into 2 gamma rays, making neutron stars
gamma-ray sources. In this letter, we strengthen considerably the limits on the
radius of compactification of extra-dimensions for small number $n$ of them, or
alternatively the fundamental scale of quantum gravity, considering the 
gamma-ray emission of the whole population of neutron stars sitting in the
Galactic bulge, instead of the closest member of this category. For $n=1$
the constraint on the compactification radius is $R<400\mu$m, overlapping
with the distance ($180\mu$m) at which Newtons law is directly measured.
In addition, for n=1 and n=2, the fundamental energy scale of quantum
gravity is far beyond the collider technology. These results imply that
if $n\lesssim4$ and if strong gravity is around a TeV, the compactification
topology is to be more complex than that of a torus.
\end{abstract}
\vspace{1truecm}
\maketitle

\section{Introduction}
Gravitation is exceedingly weaker than the electroweak interaction and,
 accordingly, the gravity energy scale, or Planck mass,
$M_{\rm Pl}=1.22\times10^{19}$~ GeV, is enormously larger than the
electroweak scale ($\approx 1$~ TeV). The extreme weakness 
of gravity relative to other interactions is a deep concern for fundamental 
physics and calls for an explanation. One elegant way to solve this hierarchy 
problem is to invoke compactified extra dimensions only allowed to gravity 
and forbidden to all other forces and particles~\cite{3,4}. Since
gravity dilutes in an extra volume, a large number of extra dimensions
or a large compactification 
radius, or both, could lower dramatically the value of the fundamental Planck 
scale without entering into contradiction with measurements of the gravity 
law at small scale. Thus, gravity would become strong at energy much lower 
than the Planck one, possibly as low as $1$ TeV. In this case, a rich 
phenomenology is expected at the Large Hadron Collider (LHC) energy, or in 
high-energy cosmic rays implying, for instance, the production of minuscule 
black holes with distinctive signatures~\cite{2}. This sounds
fascinating, but the 
duty of the physicist is to confirm or disprove this theoretical 
construction through experiments and/or observations.

In the scenario of Arkani-Hamed, Dimopoulos and Dvali~\cite{3,4}
(herein referred as ADD) with n large extra dimensions, gravity
propagates in the 4 + n 
dimensional bulk of space-time while gauge and matter fields are confined 
to a four dimensional subspace. It is assumed, for the sake of simplicity, 
that the n extra dimensions are compactified on a torus whose volume $V_n$
is expressed in terms of a common radius $R$ via $V_n\equiv(2\pi R)^n$.
Kaluza Klein (KK) gravitons which propagate in the extra 
dimensions with momentum p, will appear in our 4--D world as particles of
mass $mc = p$. The KK modes are discrete with a density of states $R^n$. The 
most intriguing feature of ADD extra dimensions is the possible production 
of KK gravitons (KKG) by nucleon-nucleon bremsstrahlung in the course of 
core collapse of massive stars~\cite{3,4,nn}. Furthermore,
Hannestad and Raffelt (herein referred as HR) pointed out the existence
of a swarm of long lived KKG around neutron stars (NS)~\cite{5} .
In the core collapse of massive stars reaching a maximum temperature
$T$, KKG of mass $m \approx T$ are copiously produced and a 
significant fraction of them remains trapped around the newly born NS. 
They slowly decay into $2 \gamma$  and e$^+$ e$^-$ pairs, making NS 
gamma-ray sources.

HR have considered various kinds of astrophysical tests implying 
high-energy radiation: gamma-ray emission of nearby NS, heating of 
the surface of NS, extragalactic MeV gamma-ray background due to the 
combination of all core collapse supernovae in the universe integrated 
over redshift. As a result, they have set the most severe constraints 
of all physics on the number n and size $R$ of extra dimensions for 
small $n$, at least in the ADD framework, much more stringent than 
the limits derived by indirect signals of extra dimensions at 
colliders~\cite{6,7} for $n = 2$ and $n = 3$. For $n \gtrsim 4$,
colliders and other 
constraints become stronger~\cite{8}. In the following, we strengthen 
considerably the limits on the radius of compactification of extra 
dimensions for small $n$, or alternatively the fundamental scale of 
quantum gravity in the ADD scheme, considering the gamma-ray emission 
of the whole population of NS sitting in the Galactic bulge (GB), 
instead of the closest member of this category. Following HR, we 
have taken a fiducial temperature $T = 30$ MeV and density
$\rho=3\times10^{14}\,{\rm g}{\rm cm}^{-3}$. For $n < 4$, the mean
mass $m$ of core collapse induced KKG is less than 90 MeV~\cite{5}, and 
thus their lifetime against the $2\gamma$ emission is $\tau_{2\gamma} = 
6 \times 10^9 (100 \mbox{MeV/m})^3$ yr, sufficiently long to neglect 
the disappearance of KKG by exponential 
decay since their production even in the oldest NS of the Galaxy.

\section{Constraints from $\gamma$-ray emission from the Galactic Bulge}
To set a limit on the high-energy gamma-ray emission from the GB, 
we use the observations of the diffuse gamma-ray emission from the 
Galaxy performed by the EGRET instrument aboard the {\it Compton 
Gamma-Ray Observatory}~\cite{9}. We concentrate in particular on the latitude 
profile of the observed flux averaged over the 
$-10^{\circ} < l < 10^{\circ}$ longitude 
interval in the 100-300 MeV energy band. Although the latitude extent 
of such a diffuse emission is fairly accurately reproduced by a model 
calculation of the emission based on a dynamical balance between the 
cosmic rays, magnetic fields, and gravitational attraction of the 
interstellar matter in the Galaxy~\cite{10}, the model prediction falls 
off faster than the observation at $|b| > 3^{\circ}$. Even if such an excess 
flux could be attributed to an under prediction of the inverse Compton 
contribution at these latitudes~\cite{9}, a fraction of it could eventually 
due to the GB. Considering that the GB extends over a $6.5^\circ$ 
radius circle around the Galactic centre, a conservative upper limit 
$F_{100-300}$ of the GB gamma-ray flux in the 100--300 MeV energy band 
can then be estimated by summing the whole excess flux which manifests 
over the $-6.5^{\circ} < b < 6.5^{\circ}$ latitude interval. 
We find $F_{100-300}=8\times 10^{-7}$
photons~
cm$^{-2}$ s$^{-1}$. Since the KKG decay emission spectrum is such that 
very few photons of energy $> 300$ MeV are produced, $F_{>   100} \approx F_{100-300}$, 
and we can set to a possible KKG induced gamma-ray flux above 100 MeV 
from the GB an upper limit $F_{>   100}=8\times 10^{-7}$ photons~ cm$^{-2}$ s$^{-1}$.

\begin{table*}[t]
\caption{\label{tab} Upper limits on the compactification radius $R$
(in m) and corresponding lower limits on the fundamental energy scale $\bar
M_{4+n}$ (in TeV), for $N_{NS}=7 \times 10^8$ and T=30 MeV. 
We show for comparison, in parentheses, the limits 
obtained by Hannestad \& Raffelt in Ref.~\cite{5}, as they would be
obtained for a distance to the NS of 0.12 kpc.}
\begin{ruledtabular}
\begin{tabular}{cccccccc}
n&1&2&3&4&5&6&7\\
\hline \\
$R$&$3.9\times10^{-4}$&$3.8\times10^{-10}$&$4.2\times10^{-12}$&
$4.7\times10^{-13}$&$1.3\times10^{-13}$&$5.4\times10^{-14}$&
$2.9\times10^{-14}$\\
&(7.7)&$(5.3\times10^{-8})$&$(1.1\times10^{-10})$&
$(5.5\times10^{-12})$&$(9.2\times10^{-13})$&$(2.8\times10^{-13})$&
$(1.2\times10^{-13})$\\
\hline \\
$\bar M_{4+n}$&$7.8\times10^{4}$&$4.5\times10^{2}$&$1.9\times10^{1}$&2.2&$4.7\times10^{-1}$&$1.47\times10^{-1}$&$5.9\times10^{-2}$\\
&$(2.9\times10^{3})$&($3.8\times10^{1}$)&(2.6)&($4.3\times10^{-1}$)&($1.2\times10^{-1}$)&($4.3\times10^{-2}$)&($1.9\times10^{-2}$)\\
\hline 
\end{tabular}
\end{ruledtabular}
\end{table*}

We aim at comparing the collective gamma-ray emission of the population 
of NS gathered in the GB to such an upper limit. The GB is a compact 
stellar system as old as globular clusters, i.e. $13 \pm 2.5$ Gyr~\cite{11}.
 The dominant stellar population of the GB is old with a broad 
metallicity distribution whose mean value is roughly solar~\cite{12,13,14}. 
There is an excess of certain $\alpha$-nuclei (as Mg) which is a clear 
signature of dominance of SNII 15 (core collapse supernovae which 
are the producers of NS). From infrared imagery, stellar dynamics 
and microlensing studies, the total GB stellar mass inferred is 
$\sim 2 \times 10^{10} M_\odot$~\cite{16}.
 Motivated by gravitational microlensing studies, the 
knowledge of the GB has increased dramatically~\cite{17}. The stellar mass 
function in the GB has been determined for low mass stars ($M < 1 M_\odot$) 
still present. Extrapolating to higher masses with a power law mass 
spectrum with an index $\alpha = -2$, Gould~\cite{18} obtains a total number of 
NS in the GB, $N_{NS} = 10^9$. This index value is consistent with the GB 
model in Ref.~\cite{19}, but differs from that adopted in Ref.~\cite{20} 
($\alpha = -2.3$)
 which is close to the conventional Salpeter one which applies to the
 disc of the Galaxy~\cite{21}. With $\alpha = -2.3$, the total number of NS in the
 GB is $N_{NS} = 4.2 \times 10^8$; a steeper index is excluded since it would lead
 to insufficient iron abundance in GB stars. In what follows, the 
total number of NS in the GB is taken to be $N_{NS} = (7 \pm 3) \times 10^8$.
The integral gamma-ray flux $F(>E_0)$ above the energy $E_0$ from a single
 NS at distance $d$ depends on the number $n$ and size $R$ of the 
extra dimensions through the expression
\begin{eqnarray}
F(>E_0)&=&8.1\times10^{-23}\,{\rm cm}^{-2}{\rm s}^{-1}
\left(\frac{d}{{\rm kpc}}\right)^{-2}\\
&&\times\left(\frac{T}{30\,{\rm MeV}}\right)^{11/2}
\left(\frac{\rho}{3\times10^{14}\,{\rm g}{\rm cm}^{-3}}\right)\nonumber\\
&&\times\Omega_n(RT)^n I_n(2E_0/T)\,,\nonumber
\end{eqnarray}
where $\Omega_n$ is the surface of the n-dimensional unit sphere, and 
$I_n(2E_0/T)$ is an integral tabulated in HR. The total expected gamma-ray 
flux above 100 MeV from the GB is obtained taking d = 8 kpc and 
multiplying by the total number of NS in the GB. Compared to the 
above derived flux limit $F_{> 100}=8\times 10^{-7}$
photons~ cm$^{-2}$ s$^{-1}$, this yields new upper limits on the
compactification volume $V_n=(2\pi R)^n$ of the compact extradimensional
torus, or equivalently lower limits on the fundamental scale of the
theory, or effective Planck scale $\bar M_{4+n}$, as defined by
$M_{\rm Pl}^2/(8\pi)=(2\pi R)^n\bar M_{4+n}^{2+n}$ in the notation of HR.

Assuming $N_{NS}=7 \times 10^8$ and T=30 MeV, 
we obtain the limits presented in Tab.~1, for $n$ ranging from 1 
to 7. Compared to a single NS located at 0.12 kpc, with an upper limit
on the gamma-ray flux above 100 MeV of $10^{-7}$
photons~ cm$^{-2}$ s$^{-1}$~\cite{5}, our limit based leads 
to a substantial gain. The very large number of NS strongly 
over-compensates the increase in distance and flux limit. 

The limits presented in Tab.~1 depend on the total number
of neutron stars $N_{NS}$ and on the temperature T. To obtain the 
limit on R for a generic value of $N_{NS}$, one should 
divide the limit in Tab.~1 by a factor 
$[N_{NS}/(7 \times 10^8)]^{1/n}$. Varying $N_{NS}$ in the
interval $3 \times 10^8 < N_{NS} < 10^9$, leads then to a 
variation of the limit on R of order of ~$\approx$40\% for n=1, 
and ~$\approx$10\% for n=7. For what concerns the dependence on the 
temperature, reducing T to 20 MeV degrades the limit on R
by a factor of about 10, 3, 2, for $n =$ 1, 2, 3, respectively. 
Conversely, increasing T to 50 MeV strengthens the limits by a 
factor of 16, 4, 2.5.

\section{Conclusions}

In conclusion, we considered the production and decay of KKG around NS
in the scenario of ADD extra dimensions and discussed the collective 
gamma-ray emission of the population of NS gathered in the GB. 
Comparing such emission to the upper limit set by EGRET observations 
of the Galaxy in the $100$~MeV$<$E$<300$~MeV range, we were able to 
constrain the compactification radius $R$ and the fundamental energy scale 
$\bar M_{4+n}$. We emphasize that for n = 1, the compactification 
radius is $<4\times 10^{-4}$ m, overlapping with the distance ($180 \mu$m) at 
which Newtons law is directly measured~\cite{22}. In addition, for 
$n = 1$ and $n = 2$, the effective Planck scale  $\bar M_{4+n}$
is far beyond the collider technology. From $n \gtrsim 4$, we recall that 
the collider and cosmic-ray constraints are more stringent than 
the astrophysical ones. But for $n < 4$, the improvement is spectacular 
as far as gamma rays are concerned. It should be noted
that comparing the observed luminosity of PSR J0953+0755~\cite{5}
with the one predicted from decaying KK modes leads to a somewhat
more stringent constraint than the ones discussed here.
However, our new constraints are more reliable since 
(1) the heating of NS, and even the notion of surface temperature 
is unclear~\cite{23} (2) our new constraints rely on the large statistical 
weight of a huge collections of objects rather than on just one
object. These results imply that 
if $n\lesssim4$ and if strong gravity is around a TeV, the compactification 
topology is to be more complex than that of a torus. Since really the
volume of the extra dimensions is constrained, the limits are
independent of the relative sizes of different dimensions, except if
one or more compactification radii $R_i$ are so small that they
cannot be excited at the astrophysical 100 MeV scale,
$R_i^{-1}\gtrsim100\,$MeV~\cite{24}. In the latter case, however,
the problem effectively becomes lower-dimensional where we have seen
that the constraints are even stronger. As a curiosity we note that
barring these caveats, unification around a TeV requires $n\geq5$,
close to the values $n=6,7$ motivated by string theory.

A similar work need to be done in the warped extra dimension model~\cite{25}.

\paragraph{Acknowledgements.} We thank Elisabeth Vangioni-Flam, Keith Olive, and Giovanni Bignami for illuminating discussions on Galactic evolution, high-energy physics and astrophysics.


\begin{thebibliography}{99}

\bibitem{3} Arkani-Hamed, N., Dimopoulos, S. \& Dvali, G. The Hierarchy Problem and New Dimensions at a Millimeter. Phys. Lett. B 429, 263-272 (1998).

\bibitem{4} Arkani-Hamed, N., Dimopoulos, S. \& Dvali, G. Phenomenology, Astrophysics, and Cosmology of Theories with Submillimeter Dimensions and TeV Scale Quantum Gravity. Phys. Rev. D 59, 086004 (1999).

\bibitem{2} Giddings, S.B. \& Katz, E. Effective Theories and Black Hole Production in Warped Compactifications. J. Math. Phys. 42, 3082-3102 (2001).

\bibitem{nn} S.~Cullen, S. \& Perelstein, M. SN1987A constraints on
large compact dimensions. Phys. Rev. Lett. 83, 268 (1999);
Hanhart, C., Phillips, D.~R., Reddy, S. \& and Savage, M.~J.
Extra dimensions, SN1987a, and nucleon nucleon scattering data.
Nucl. Phys. B 595, 335 (2001).

\bibitem{5} Hannestad, S. \& Raffelt, G.G. Supernova and Neutron-Star Limits on Large Extra Dimensions Reexamined. Phys. Rev. D 67, 125008 (2003).

\bibitem{6} Rizzo, T.G. Testing the Nature of Kaluza-Klein Excitations at Future Lepton Colliders. Phys. Rev. D 61, 055005 (2000).

\bibitem{7} Cheung, K. Collider Phenomenology for Models of Extra Dimensions. Talk given at the invited session of Beyond the Standard Model in the APS/DPF April meeting; preprint at http://xxx.lanl.gov/hep-ph/0305003 (2003)

\bibitem{8} Anchordoqui, L. \& Goldberg, H. Black Hole Chromosphere at the CERN LHC. Phys. Rev. D 67, 064010 (2003).

\bibitem{9} Hunter, S.D. et al. EGRET Observations of the Diffuse Gamma-Ray Emission from the Galactic Plane. Astrophys. J. 481, 205-240 (1997).

\bibitem{10} Bertsch, D.L. et al. Diffuse Gamma-Ray Emission in the Galactic Plane from Cosmic-Ray, Matter, and Photon Interactions. Astrophys. J. 416, 587-600 (1993).

\bibitem{11} Rich, R.M. in Astrophysical Ages and Timescales (eds von Hippel, T., Mansuset, N. \& Simpson, C.) 216 (ASP Conference Series 245, Astronomical Society of the Pacific, San Francisco, 2001); preprint at http://xxx.lanl.gov/astro-ph/0108107.

\bibitem{12} McWilliam, A. \& Rich, R.M. The First Detailed Abundance Analysis of Galactic bulge K Giants in Baade's Window. Astrophys. J. Suppl. Ser. 91, 749-791 (1994).

\bibitem{13} Rich, R.M. \& McWilliam, A. in Discoveries and Research Prospects from 8- to 10-Meter-Class Telescopes (ed. Bergeron J.) 150-161 (SPIE 4005, 2000); preprint at http://xxx.lanl.gov/astro-ph/0005113.

\bibitem{14} Zoccalli, M. et al. Age and Metallicity Distribution of the Galactic Bulge from Extensive Optical and Near-IR Stellar Photometry. Astron. Astrophys. 399, 931-956 (2003).

\bibitem{15} Matteucci, F., Renda, A., Pipino, A. \& Della Valle, M. Modelling the Nova Rate in Galaxies. Astron. Astrophys. 405, 23-30 (2003).

\bibitem{16} Zhao, H., Rich, R.M. \& Spergel, D.N. A Consistent Microlensing Model for the Galactic Bar. Mon. Not. R. Astron. Soc. 282, 175-181 (1996).

\bibitem{17} Han, C. \& Gould, A. Stellar Contribution to the Galactic Bulge Microlensing Optical Depth. Astrophys. J. 592, 172-175 (2003).

\bibitem{18} Gould, A. Measuring the Remnant Mass Function of the Galactic Bulge. The Astrophysical Journal, Astrophys. J. 535, 928-931 (2000).

\bibitem{19} Matteucci, F. Romano, D. \& Molaro, P. Light and Heavy Elements in the Galactic Bulge. Astron. Astrophys. 341, 458-468 (1999).

\bibitem{20} Nakasato, N. \& Nomoto, K. Three-Dimensional Simulations of the Chemical and Dynamical Evolution of the Galactic Bulge. Astrophys. J. 588, 842-851 (2003).

\bibitem{21} Kroupa, P. The Initial Mass Function of Stars: Evidence for Uniformity in Variable Systems. Science 295, 82-91 (2002).

\bibitem{22} Adelberger, E.G., Heckel, B.R. \& Nelson, A.E. Tests of the Gravitational Inverse-Square Law. Annu. Rev. Nucl. Part. Sci. 53, in the press (2003); preprint at <http://xxx.lanl.gov/hep-ph/0307284>.

\bibitem{23} Pavlov, G.G. \& Zavlin, V.E. in Proceedings of the XXI Texas Symposium on Relativistic Astrophysics; preprint at http://xxx.lanl.gov/astro-ph/0305435.

\bibitem{24} Lykken, J. \& Nandi, S. Asymmetrical Large Extra Dimensions. Phys. Lett. B 485, 224-230 (2000). 

\bibitem{25} Randall, L. \& Sundrum, R. Large Mass Hierarchy from a Small Extra Dimension. Phys. Rev. Lett. 83, 3370-3373 (1999).

\end{thebibliography}
\end{document}